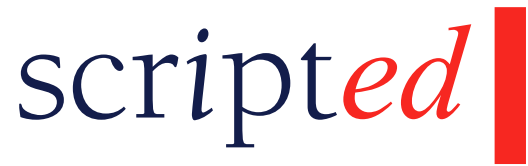

# Copyright in AI-generated works: Lessons from recent developments in patent law

*Rita Matulionyte\* and Jyh-An Lee\*\**

**Abstract**

In *Thaler v The Comptroller-General of Patents, Designs and Trade Marks (DABUS)*, Smith J held that an AI owner can possibly claim patent ownership over an AI-generated invention based on their ownership and control of the AI system. This AI-owner approach reveals a new option to allocate property rights over AI-generated output. While this judgment was primarily about inventorship and ownership of AI-generated invention in patent law, it has important implications for copyright law. After analysing the weaknesses of applying existing judicial approaches to copyright ownership of AI-generated works, this paper examines whether the AI-owner approach is a better option for determining copyright ownership of AI-generated works. The paper argues that while contracts can be used to work around the AI-owner approach in scenarios where users want to commercially exploit the outputs, this approach still provides more certainty and less transaction costs for relevant parties than other approaches proposed so far.

**Keywords**

artificial intelligence, computer-generated work, AI-generated work, DABUS, patent, copyright, ownership

\* Senior Lecturer, Macquarie Law School, Sydney, Australia, rita.matulionyte@mq.edu.au.
\*\* Professor, Faculty of Law, The Chinese University of Hong Kong, Hong Kong SAR.

An earlier version of the paper was presented in the *13th IP Conference – Innovation, Intangible Assets During and After the Global Pandemic* held by the Chinese University of Hong Kong Faculty of Law. The autors thank Albert Wai-Kit Chan, Peter Yu and anonymous reviewers for their helpful comments.

# 1  Introduction

Artificial Intelligence (AI) technologies, such as machine learning (ML), are used widely in both the public and private sectors in various applications, such as online advertising, medical research and diagnosis, and facial recognition.[1] AI/ML technologies are also being increasingly adopted by creative industries to generate outputs that would normally be protected under copyright law, such as art,[2] music,[3] novels,[4] and even film scripts.[5] Some of these creative outputs have been successfully commercialised, such as the Portrait of Edmund Belamy, an AI-generated work that was sold at a Christie's auction for $432,500.[6]

---

[1] E.g. Zeynep Tufekci, "How Recommendation Algorithms Run the World" (*Wired*, 22 April 2019), available at https://www.wired.com/story/how-recommendation-algorithms-run-the-world/ (accessed 20 April 2021); Hafizah Osman, "New AI Tech Reshapes Skin Cancer Detection" (*Healthcareit*, 30 January 2019), available at https://www.healthcareit.com.au/article/new-ai-tech-reshapes-skin-cancer-detection (accessed 20 April 2021); Yason Tashea, "Courts Are Using AI to Sentence Criminals. That Must Stop Now" (*Wired*, 17 April 2017), available at https://www.wired.com/2017/04/courts-using-ai-sentence-criminals-must-stop-now/ (accessed 20 April 2021); Asha Barbaschow, "AFP used Clearview AI Facial Recognition Software to Counter Child Exploitation" (ZDnet, 15 April 2020), available at https://www.zdnet.com/article/afp-used-clearview-ai-facial-recognition-software-to-counter-child-exploitation/ (accessed 20 April 2021).

[2] Gabe Cohn, "AI Art at Christie's Sells for $432,500" (*The New York Times*, 25 October 2018), available at https://www.nytimes.com/2018/10/25/arts/design/ai-art-sold-christies.html (accessed 20 April 2021).

[3] "Warner Music Signs First Ever Record Deal with an Algorithm", (*The Guardian*, 23 March 2019), available at https://www.theguardian.com/music/2019/mar/22/algorithm-endel-signs-warner-music-first-ever-record-deal (accessed 20 April 2021).

[4] Chloe Olewitz, "A Japanese A.I. Program Just Wrote a Short Novel, and it Almost Won a Literary Prize" (*Digital Trends*, 23 March 2016), available at https://www.digitaltrends.com/cool-tech/japanese-ai-writes-novel-passes-first-round-nationanl-literary-prize/ (accessed 20 April 2021).

[5] Annalie Newitz, "Movie Written by Algorithm Turns Out to Be Hilarious and Intense" (*ArsTechnica*, 6 September 2016), available at https://arstechnica.com/gaming/2016/06/an-ai-wrote-this-movie-and-its-strangely-moving/ (accessed 20 April 2021).

[6] Cohn, *supra* n. 2.

In the last few years, legal professionals, academics,[7] and policy bodies around the world[8] have been actively discussing whether AI-generated works are, and should be, protected under copyright laws and, if so, who should own the copyright. Except for a few jurisdictions that have provisions on computer-generated works (e.g., the UK, Ireland, New Zealand, India, and Hong Kong) in their copyright laws,[9] AI-generated works are not easily copyrighted in most countries because such laws require human authorship for copyright protection.[10] At the policy level, there is no consensus on whether copyright law protection should be extended to AI-generated works. While some suggest that works autonomously generated by AI (as opposed to AI-assisted works) do not need copyright protection,[11] others argue that granting copyright protection for such works would increase incentives to develop sophisticated AI technology and eventually lead to more original creations reaching the public.[12] Many who believe protection is desirable have considered different ownership allocation options, such as allocating authorship and initial ownership to the coder/developer of the AI, the user of the AI, or even to the AI itself.[13] However, no consensus has been reached on which option is the most suitable.

---

[7] E.g. Courtney White and Rita Matulionyte, "Artificial Intelligence Painting a Larger Picture on Copyright" (2020) 30(4) *Australian Intellectual Property Review* 224-242; Russ Pearlman, "Recognizing Artificial Intelligence (AI) As Authors and Inventors Under U.S. Intellectual Property Law" (2018) 24(2) *Richmond Journal of Law and Technology* 1-38; Ana Ramalho, "Will Robots Rule The (Artistic) World? A Proposed Model For The Legal Status Of Creations By Artificial Intelligence Systems" (2017) 21(1) *Journal of Internet Law* 12-25; Rex M. Shoyama, "Intelligent Agents: Authors, Makers, and Owners of Computer-Generated works in Canadian Copyright Law" (2005) 4(2) *Canadian Journal of Law and Technology* 129-140; Julia Dickenson, Alex Morgan, and Birgit Clark, "Creative Machines: Ownership of Copyright in Content Created by Artificial Intelligence Applications" (2017) 39 *European Intellectual Property Review* 457-460, pp. 457-458; Tim W Dornis, "Artificial Creativity: Emergent Works and the Void in Current Copyright Doctrine" (2020) 22 *Yale Journal of Law & Technology* 1-60, pp. 20-24; Andres Guadamuz, "Do Androids Dream of Electric Copyright? Comparative Analysis of Originality in Artificial Intelligence Generated Works" (2017) 2 *Intellectual Property Quarterly* 169-186, pp. 182-183; Amir H Khoury, "Intellectual Property Rights for 'Hubots': On the Legal Implications of Human-Like Robots as Innovators and Creators" (2017) 35(3) *Cardozo Arts and Entertainment Law Journal* 635-668; Massimo Maggiore, "Artificial Intelligence, Computer Generated Works and Copyright" in Enrico Bonadio and Nicola Lucchi (eds*.) Non-Conventional Copyright: Do New and Atypical Works Deserve Protection?* (Cheltenham: Edward Elgar, 2018), pp. 387-389.

[8] E.g. "UK Government Consultation on Artificial Intelligence and Intellectual Property", available at https://www.gov.uk/government/consultations/artificial-intelligence-and-intellectual-property-call-for-views (accessed 20 April 2021); European Parliament resolution of 20 October 2020 on intellectual property rights for the development of artificial intelligence technologies (2020/2015(INI)); USPTO, "Public Views on Artificial Intelligence and Intellectual Property" (October 2020), available at https://www.uspto.gov/sites/default/files/documents/USPTO_AI-Report_2020-10-07.pdf (accessed 20 April 2021); The WIPO Conversation on Artificial Intelligence and Intellectual Property, available at www.wipo.org (accessed 20 April 2021).

[9] See Copyright, Design and Patents Act 1988 (UK), s. 9(3); Copyright Act 1994 (NZ), s. 5(2)(a); Copyright Act 1957 (India), s. (2)(d)(vi); Copyright Ordinance (Hong Kong) cap 528, s. 11(3); Copyright and Related Rights Act 2000 (Ireland), s. 21(f).

Similarly, there have been discussions on whether AI-generated outputs should be protected under patent law, who should be considered the inventor, and who should own an AI-generated invention. While some have suggested that AI-generated outputs should be protected by patent law,[14] others have expressed serious concerns about the impacts that patent protection over AI-generated outputs could have.[15] Those who agree that patent protection should be awarded in such situations have proposed suggestions on the allocation of inventorship and initial ownership, such as allocating initial ownership to the AI itself,[16] to the owner of the AI system (i.e., 'computer' or 'machine'),[17] to the user of the AI system,[18] or to the hardware or software developer.[19]

---

[10] E.g. Jani Ihalainen "Computer Creativity: Artificial Intelligence and Copyright" (2018) 13(9) *Journal of Intellectual Property Law & Practice* 724-728, pp. 726-727; Paul Lambert, "Computer-Generated Works and Copyright: Selfies, Traps, Robots, AI and Machine Learning" (2017) 39(1) *European Intellectual Property Review* 12-20, p. 14; Maggiore, *supra* n. 7, pp. 387-389; Mark Perry and Thomas Marhoni, "From Music Tracks to Google Maps: Who Owns Computer-Generated Works?" (2010) 26(6) *Computer Law & Security Review* 621-629, pp. 624-625; Ramalho, *supra* n. 7, pp. 14-16; Jacob Turner, *Robot Rules Regulating Artificial Intelligence* (Basingstoke: Palgrave Macmillan, 2019), pp. 123-124.

[11] E.g. Patrick Zurth, "Artificial Creativity? A Case Against Copyright Protection for AI Generated Works", *UCLA Journal of Law & Technology* (forthcoming).

[12] E.g. Shlomit Yanisky-Ravid and Luis Antonio Velez-Hernandez, "Copyrightability of Artworks Produced by Creative Robots, Driven by Artificial Intelligence Systems and the Originality Requirement: The Formality-Objective Model" (2018) 19(1) *Minnesota Journal of Law, Science & Technology* 1-54; Pearlman,*supra* n. 7; Jane C. Ginsburg and Luke A. Budiardjo, "Authors and Machines", (2019) 34(2) *Berkeley Technology Law Journal* 343-448.

[13] E.g. Shlomit and Velez-Hernandez, *supra* n. 12 (suggesting that AI should hold copyright); Pearlman, *supra* n. 7; Ginsburg and Budiardjo, *supra* n. 12 (suggesting different ownership allocation options depending on contributions).

[14] See Ryan Abbott, "I Think, Therefore I Invent: Creative Computers and the Future of Patent Law" (2016) 57(4) *Boston College Law Review* 1079-1126; Erica Fraser, "Computers as Inventors – Legal and Policy Implications of Artificial Intelligence on Patent Law" (2016) 13(3) *SCRIPTed*

The recent DABUS case concerned application for a patent wherein the AI system DABUS was listed as an inventor of the claimed invention. The application was initially lodged before the US and UK patent offices and subsequently extended to other national offices, including the European Patent Office and national patent offices in Germany, Japan, South Korea, Australia, and Israel. The patent application was rejected by the US, UK, European, and Australian patent offices, all of which concluded that an AI system is not eligible for inventorship.[20] All of these decisions were appealed, and the High Court of England and Wales (UK) was the first court to assess – and eventually reject – the appeal of the decision in *Thaler v The Comptroller-General of Patents, Designs And Trade Marks* ('*DABUS*').[21] Interestingly, while the High Court of England and

---

Wales held that, under current UK patent law, the AI machine DABUS cannot be listed as an inventor, the court recognised the possibility of listing the owner of the AI DABUS (in this case, Thaler) as both the inventor and the owner of the AI-generated invention.[22]

Although the *DABUS* case is primarily about inventorship in patent law, we argue that it also provides important implications for the debate on AI and copyright. Specifically, we consider whether Smith J's viewpoint on patent ownership allocation, as indicated above, could be applied in copyright law. Namely, we aim to determine if it is reasonable to assign ownership of AI-generated works to the AI owner. While this option has been, to some extent, discussed in commentary on patent law,[23] it has not been significantly discussed in relation to copyright law.

This article revisits the existing discussion on copyright ownership of AI-generated works and critically assesses whether allocating ownership of AI-generated works to the AI owner is a more desirable option than those proposed previously. This paper does not weigh in on the important debate over whether copyright should subsist in AI-generated works, which has been discussed by a significant portion of the literature.[24] Rather, assuming that such protection is desirable, this paper focuses on who should own the copyright of AI-generated works. Based on court decisions regarding copyright protection over computer-generated works in the UK and China, the paper first discusses the strengths and weaknesses of the most frequently proposed options for ownership of AI-generated works: the AI software developer and the AI software user. It then examines the patent ownership allocation rule proposed in *DABUS* and considers whether this rule should be applied in determining copyright for AI-generated works. The paper concludes by identifying the main advantages and issues that an AI-owner rule would pose in the domain of copyright law.

---

[22] It should be noted that while the Court of Appeals upheld the High Court's decision on 21 September 2021, the former did not specifically address whether Thaler could be listed as both the inventor and the owner of the AI-generated invention. *Thaler v The Comptroller-General of Patents, Designs And Trade Marks* [2021] EWCA Civ 1374.

[23] Abbott, *supra* n. 14; Davies, *supra* n. 16, p. 618; see also Vertinsky and Rice, *supra* n. 15, p. 609.

[24] E.g. Jyh-An Lee, "Computer-generated Works under the CDPA 1988" in Jyh-An Lee, Reto Hilty and Kung-Chung Liu (eds), *Artificial Intelligence and Intellectual Property* (Oxford: Oxford University Press, 2021), pp. 183-194; White and Matulionyte, *supra* n. 7, pp. 232-236; Burkhard Schafer at al., "A Fourth Law of Robotics? Copyright and the Law and Ethics of Machine Co-Production" (2015) 23 *Artificial Intelligence and Law* 217-240, pp. 227-230.

## 2 Current approaches to copyright ownership of AI-generated works: Software developer or user?

Those who argue that AI-generated works should be subject to copyright protection have different viewpoints regarding who should be the owner of such works. The most common proposals are that such copyright should be allocated to the AI software developer, the AI software user, or even the AI itself.[25] Discussions on the ownership of AI-generated works are becoming less hypothetical as courts in a few jurisdictions have begun to confront the issue.

In this section, we use English and Chinese cases as examples illustrating existing judicial approaches to this issue. We focus on two approaches adopted in English and Chinese case law: the software developer as the owner and the software user as the owner of AI-generated works. We do not discuss the option of allocating ownership to the AI itself, both because this approach has never been taken by any court and because AI lacks the legal personhood necessary to have legal rights.[26] Thus, from both a theoretical and practical perspective, the proposal of allocating the copyright to AI itself has been ruled out.

---

[25] Text accompanying n. 3.
[26] Annemarie Bridy, "Coding Creativity: Copyright and the Artificially Intelligent Author" (2012) 2012 *Stanford Technology Law Review* 5-28, p. 51; Eliza Mik, "AI as a Legal Person?" in Jyh-An Lee, Reto Hilty and Kung-Chung Liu (eds.), *Artificial Intelligence and Intellectual*

**2.1 The software developer as the owner**

Software developers provide an essential contribution to the creation of AI-generated works. Although these works are not directly created by software developers, they would not exist in the first place without the developer's software.[27] It stands to reason, then, that software developers have the potential for ownership of AI-generated works produced using their software. The concept of a software developer is broadly defined: It can be used to refer to an individual programmer who develops the software, or it can be used to refer to a company that hires programmers to develop the software that is subsequently owned by the company.

This section introduces the *Nova* case from the UK and the *Tencent* case from China, in which the courts ruled that the software developers were the rightful owners of the computer-generated works. In *Nova*, provisions in the Copyright, Designs and Patents Act (CDPA) 1988 were used to determine copyright ownership of the computer-generated works, and *Tencent* used originality doctrine. Both courts emphasised that the justification behind giving software developers ownership of the computer-generated outputs was that they had substantially determined how the outputs were arranged.

*2.1.1 The Nova Case in the UK*

The CDPA 1988 in the United Kingdom (UK) provides copyright protection for literary, dramatic, musical, and artistic works generated by computers under circumstances without human author.[28] In other words, for a computer-

---

generated work in the UK, human authorship is irrelevant to whether the work is copyrightable. The CDPA 1988 further stipulates that the author of the computer-generated work is 'the person by whom the arrangements necessary for the creation of the work are undertaken'.[29] Some commentators view the computer-generated work provisions in the CDPA 1988 as innovative,[30] and some believe it was the first legislation in the world protecting copyright in the context of AI.[31]

While commenters in some countries have advocated adopting the computer-generated works provisions from the CDPA 1998 to cope with new challenges raised by AI technologies,[32] the British courts have only applied these provisions once: in *Nova Productions v Mazooma Games and Others*, a case which did not involve any AI technology.[33] The work concerned was the display of a series of composite frames generated by a computer program using bitmap files in a coin-operated game 'Pocket Money' that was designed, manufactured, and sold by the claimant Nova Productions Limited ('Nova').[34] Kitchin J in *Nova* considered whether the computer-generated work in a computer game belonged to the programmer or the user:

> In so far as each composite frame is a computer generated work then the arrangements necessary for the creation of the work were undertaken by [the programmer] Mr. Jones because he devised the appearance of the various

---

[29] CDPA 1988, s. 9(3).
[30] Ysolde Gendreau, "Copyright Ownership of Photographs in Anglo-American Law" (1993) 15(6) *European Intellectual Property Review* 207-211, pp. 210-211.
[31] Toby Bond and Sarah Blair, "Artificial Intelligence and Copyright: Section 9(3) or Authorship without an Author" (2019) 14(6) *Journal of Intellectual Property Law & Practice* 423.
[32] Bridy, *supra* n. 26, pp. 66-67; Cody Weyhofen, "Scaling the Meta-Mountain: Deep Reinforcement Learning Algorithms and the Computer-Authorship Debate" (2019) 87(4) *UMKC Law Review* 979-996, p. 996.
[33] *Nova Productions Ltd v Mazooma Games Ltd* [2006] EWHC 24 (Ch) (20 January 2006).
[34] *Ibid.*, paras. 12-18.

> elements of the game and the rules and logic by which each frame is generated and he wrote the relevant computer program. In these circumstances I am satisfied that Mr. Jones is the person by whom the arrangements necessary for the creation of the works were undertaken and therefore is deemed to be the author by virtue of s.9(3).[35]

As for the role of the player/user in the game, Kitchin J ruled the following:

> The appearance of any particular screen depends to some extent on the way the game is being played. For example, when the rotary knob is turned the cue rotates around the cue ball. Similarly, the power of the shot is affected by the precise moment the player chooses to press the play button. The player is not, however, an author of any of the artistic works created in the successive frame images. His input is not artistic in nature and he has contributed no skill or labour of an artistic kind. Nor has he undertaken any of the arrangements necessary for the creation of the frame images. All he has done is to play the game.[36]

While Kitchin J's analysis in *Nova* seems plausible in determining copyright ownership between the programmer and player in the video game, allocating copyright to the programmer instead of the user of the computer-generated work is not always self-evident in all applications of software technologies and, in particular, AI technologies. First, AI algorithms are different from traditional software, as the former requires a huge volume of data with which to train the machine. Because there are other equally important stakeholders, such as trainers and data providers, involved in the development of the AI software, programmers are not the only party that enable the operation of an AI application. Second, while software developers provide step-by-step instructions for the machine to follow in traditional computer programming, AI algorithms function through the observation of data instead of encoded instructions.[37] Therefore, software developers have much less control over how a work is generated by the algorithm in the AI environment than in traditional computer programming. Consequently, the legal treatment of a software developer as determined in *Nova* might need to be adjusted based on AI's technical character. Last, but not least, there are many scenarios other than video games where the works are generated because of users' operation of the software. If users generate commercially valuable content for their own business purposes, they will certainly have more interest in using the content than video game players and software developers.[38] Therefore, assigning copyright of the AI-generated works to the software developer is not always straightforward.

---

[35] *Ibid.*, paras. 105-106.
[36] *Ibid.*

### 2.1.2 Tencent case in China

Most jurisdictions do not have computer-generated work provisions in their copyright laws like the UK and a few other commonwealth jurisdictions do.[39] For example, in the United States and most European countries, AI-generated works are not copyrightable because of the absence of human creativity in their creation.[40] Thus, it is challenging for software developers to claim ownership

---

[37] Megan Sword, "To Err Is Both Human and Non-Human" (2019) 88(1) *UMKC Law Review* 211-233, p. 213.
[38] E.g. the *Feilin* case in section 2.2.
[39] It should be noted that not all Commonwealth jurisdictions have computer-generated work clauses similar to those in the CDPA 1988. Australia is a notable example; the courts ruled that computer-generated works were not copyrightable because there was no human author and the works thus lacked originality – see IceTV [2009] HCA 14.
[40] Enrico Bonadio, Luke McDonagh, and Christopher Arvidsson, "Intellectual Property Aspects of Robotics" (2018) 9(4) *European Journal of Risk Regulation* 655-676, p. 669; Jeremy A. Cubert and Richard G.A. Bone, "The Law of Intellectual Property Created by Artificial Intelligence" in Woodrow Barfield and Ugo Pagallo (eds.) *Research Handbook on the Law of Artificial*

over works autonomously generated by AI. However, in the recent Chinese case *Tencent v Shanghai Yingxun Technology Co. Ltd*, Tencent successfully convinced the court that the software developer contributed originality to an AI-generated work and, therefore, should be its owner.[41] From a comparative law perspective, this is an exceptional case, as it is unusual for the court to rule that AI developers exercised skill and judgment in an AI-generated work.

The disputed work in *Tencent* was an article about the Shanghai stock market written by the plaintiff's AI software Dreamwriter.[42] Dreamwriter collected data from multiple sources, analysed the data using its machine-learning algorithms, verified the data, wrote an article using the verified data, and then published it.[43] The defendant argued that the article was not copyrightable because there was no human creativity involved in its production.[44] However, the court was convinced by the plaintiff that human originality could be found in different phases of Dreamwriter's process of creating the article. The court explained that, although it only took Dreamwriter two minutes to produce the disputed article which was the result of the software's operation of established rules, algorithms, and templates without any

---

*Intelligence* (Cheltenham: Edward Elgar, 2018), pp. 424-425; Madeleine de Cock Buning, "Autonomous Intelligent Systems as Creative Agents under the EU Framework for Intellectual Property" (2016) 7(2) *European Journal of Risk Regulation* 310-322, pp. 314-315; Julia Dickenson, Alex Morgan, and Birgit Clark, "Creative Machines: Ownership of Copyright in Content Created by Artificial Intelligence Applications" (2017) 39(8) *European Intellectual Property Review* 457-460, pp. 457-458; Dornis, *supra* n. 7, pp. 20-24; Guadamuz, *supra* n. 7, pp. 182-183; Ihalainen, *supra* n. 10, pp. 726-727; Lambert, *supra* n. 10, p. 14; Maggiore, *supra* n. 7, pp. 387-389; Perry and Marhoni, *supra* n. 10, pp. 624-625; Ramalho, *supra* n. 7, pp. 14-16; Turner, *supra* n. 10, pp. 123-124.

[41] *Tencent v. Shanghai Yingxun Technology Co. Ltd*, People's Court of Nanshan (District of Shenzhen) (2019) Yue 0305 Min Chu No. 14010 (深圳市南山区人民法院(2019)粤 0305 民初 14010 号民事判决), 24 December 2019.

[42] *Ibid*.

[43] *Ibid*.

[44] *Ibid*.

human participation, the automatic operation of Dreamwriter did not occur without a reason.[45] They also noted that the software was not self-aware.[46] Instead, Dreamwriter's autonomous operation reflected its developers' personalised selection and arrangement of data type, data format, the conditions that triggered the writing of the article, the templates of article structure, the setting of the corpus, and the training of the intelligent verification algorithm model.[47] The court in *Tencent* viewed the software developer as the owner of the AI-generated work based on originality doctrine in copyright law. The way that the court determined originality was similar to that typically applied in cases involving compilation, which was that 'the selection or arrangement of…[existing] contents constitute intellectual creations'.[48] The court determined that originality existed in the developer's choices in setting the criteria for the selection and arrangement of existing data, which was subsequently used by the AI to complete the selection and arrangement.[49]

### 2.2   Software user as the owner – the *Feilin* case in China

While the plaintiff's strategy in *Tencent* for proving the developer's contribution in the AI-generated work was successful in that litigation, the court's finding of originality is not applicable to all AI creations. Because of their nested non-linear structure, AI models are usually applied in a black-box manner. Therefore, their 'interpretability' or 'explainability' – that is, the degree to which a human observer can intrinsically understand the cause of a decision by the system – has drawn significant attention in recent years.[50] Sometimes even AI developers are unable to fully understand AIs' decision-making process or predict the systems decisions or outputs.[51] Thus, there are flaws in the argument that all AI works are well designed and that their products can be anticipated by their developers. In other words, it is conceivable that not all parts of an AI work reflect the developer's skill or judgment and, hence, the finding of originality in *Tencent* is not universally applicable. Moreove, sometimes the involvements of other parties, such as machie operators, trainers, and data providers, are essential in the production AI-generated works.[52] Many AI developers are not able to substantially envisage the AI-genearted works because they cannot control or plan other parties' data provision or processing behaviours.[53] The role of these developers is much more marginal than those in *Nova* and *Tencent* in the production of computer-genarted works. This difference also reveals that the software developer as the owner approach is not an universally justified and ideal option.

---

[45] *Ibid*.
[46] *Ibid*.
[47] *Ibid*.
[48] The Agreement on Trade-Related Aspects of Intellectual Property Rights (TRIPS), Art. 10; WIPO Copyright Treaty, Art. 5.
[49] *Tencent v. Shanghai Yingxun Technology Co. Ltd*, *supra* n. 41.

Not all courts held that software developers are justified to be the owners of computer-generated works. Some have argued that software's users are the appropriate owners of AI-generated works because they provide considerable inputs into shaping the outputs.[54] Also, a software user might be more economically affected by the ownership allocation of AI-generated works than the software developer because the former deploys the AI software to produce output for his or her own commercial interest.

---

[50] Yavar Bathaee, "The Artificial Intelligence Black Box and the Failure of Intent and Causation" (2018) 31(2) *Harvard Journal of Law & Technology* 889-938, pp. 901-906; Ashley Deeks, "The Judicial Demand for Explainable Artificial Intelligence" (2019) 119(7) *Columbia Law Review* 1829-1850, pp. 1832-1838.

[51] Nadia Banteka, "Artificially Intelligent Persons" (2021) 58(3) *Houston Law Review* 537-596, pp. 547-548; Jonathan A Schnader, "Mal-Who? Mal-What? Wal-Where? The Future Cyber-Threat of A Non-Fiction Neuromance: Legally Un-attributable, Cyberspace-Bound, Decentralized Autonomous Entities" (2019) 21(2) *North Carolina Journal of Law & Technology* 1-40, p. 34.

[52] Lee, *supra* n. 24, p. 192.

[53] Whita and Matulionyte, *supra* n. 7, p. 238.

[54] Yu, *supra* n. 27, p. 1259.

In the recent Chinese case *Feilin v Baidu*, which was decided by the Beijing Internet Court (BIC), the disputed work was an article titled "Judicial Big Data in the Film, Television and Entertainment Industry" published by the plaintiff.[55] The defendant argued that the article was not copyrightable because it was purely the result of the plaintiff's search in the Wolters Kluwer legal database.[56] The result was presented by the Wolters Kluwer Database as an analytical report, which included statistics and corresponding charts on types of claims, procedures, industries involved, amount of the claims, decision-making time, courts, judges, lawyers and firms, and frequently cited statutes in court decisions concerning the entertainment industry.[57]

The court eventually ruled for the plaintiff because the latter created original content other than the search result in the disputed article; however, the court also shed light on the ownership issue with regard to the results of the search using the Wolters Kluwer Database.[58] The court explained that there were two key players involved in the process: the programmer who developed the database software and the user who used the database to produce the search results.[59] They determined that neither the programmer nor the user could be the author of the search result: The programmer did not search in the database by imputing keywords, and thus the search result was not a reflection of his original expression,[60] and the user only typed in the keywords used to search the database, which was not an original expression under copyright law either. Thus, the search result was created by the Wolters Kluwer Database based on the input keywords, algorithms, rules, and models.[61] However, Wolters Kluwer Database was not an author because it is not considered a natural person under the law.[62]

---

[55] *Feilin v Baidu*, Beijing Internet Court, (2018) Jing 0491 Min Chu No. 239 (北京互联网法院 (2018) 京 0491 民初 239 号民事判决), 26 April 2019.
[56] *Ibid*.
[57] *Ibid*.
[58] *Ibid*.
[59] *Ibid*.
[60] *Ibid*.

Interestingly, the court went beyond the existing law to analyse policy issues regarding the legal rights over the search result. Recognising the commercial and communicative value of computer-generated works, the court indicated that allocating certain rights over the works to private parties was better than leaving them in the public domain.[63] Between the software developer and user, the court determined that it was the latter that deserved legal protection.[64] The argument was, first, that the developer had already recouped their investment in developing the software via a licensing fee or ownership of intellectual property rights into software.[65] Second, compared to the software developer, the software user had more incentive to use and disseminate the computer-generated works because they had typed in the keywords to initiate the search and had a plan for the use of the works.[66] Thus, assigning rights to the computer-generated works to the user rather than the software developer would better foster cultural and scientific development, as the user had substantive incentive to use and disseminate the works.[67]

The above reasoning of the BIC was not made using the existing Chinese copyright law, and it was not the primary conclusion of the judgment. It was, at most, the judge's personal normative viewpoint. Nevertheless, this reasoning presents a different position in favour of the software user rather than the developer regarding ownership of computer-generated and/or AI-generated works. While the BIC was correct that the software user in this case had greater interest in using the resulting works than the software developer, the assignment of relevant rights to software users has the same problem as the conclusion of the *Nova* rule, which was to assign the ownership to the software developer. As exemplified in *Nova* and *Feilin*, software users' interests in the resulting works vary from case to case. Although the user in *Feilin* had more substantial interests in utilising the resulting works than the software developers, not all users of AI algorithm or software have similar interests.[68] Moreover, while some users contribute significantly to the AI-generated works, others' contribution is negligible.[69] Thus, neither the reasoning in *Feilin* nor that of *Nova* can be the singular determinant of the optimal solution in all cases involving AI-generated works. Moreover, in *Feilin*, the user only typed in "film" as the search keyword, and the analytical report was automatically produced by the Wolters Kluwer Database.[70] Given the user's negligible contribution to the resulting work and their insignificant investment in the software system, assigning an exclusive right of ownership to them might not be justified. While the user has a substantial interest in utilising the search result, a license from a more legitimate owner could serve the same function.

---

[61]  *Ibid*.
[62]  *Ibid*.
[63]  *Ibid*.
[64]  *Ibid*.
[65]  *Ibid*.
[66]  *Ibid*.
[67]  *Ibid*.

## 3   New: The AI owner as the owner of AI-generated outputs

The above analysis has shown the weaknesses of the current proposals to allocate copyright ownership of AI-generated works to either software developers or users. In this section, we will explore another option, which is to allocate copyright of such works to the AI owner, as suggested in the UK *DABUS* decision. Although *DABUS* concerns patent inventorship and ownership, we argue that the ownership allocation rule proposed in the case also has important implications for copyright. This section will examine whether allocating ownership of AI-generated works to the AI owner would be a more viable solution than those previously discussed.

---

[68] E.g. the *Nova* case in section 2.1.1.
[69] E.g. Whita and Matulionyte, *supra* n. 7, p. 239.
[70] *Feilin v Baidu*, *supra* n. 55.

## 3.1 Why the patent law debate is relevant

Patent law and copyright law are similarly premised on the economic rationale of incentivising creativity and innovation. Thus, legal doctrines from these two fields often influence each other. For example, in *Metro-Goldwyn-Mayer Studios, Inc. v Grokster Ltd.*, the US Supreme Court borrowed from patent law to establish liability for inducement in copyright infringement.[71] When extending the staple article of commerce doctrine from patent law to copyright law in *Sony Corp. of America v Universal City Studios, Inc.*, the US Supreme Court explained that although copyright law and patent law "are not identical twins", their similarities made patent law an appropriate source from which to borrow.[72]

Likewise, many scholars have advocated for more harmonisation of the rules governing ownership in these two fields.[73] Despite differences, patent law and copyright law share substantially similar rules on initial ownership allocation. Under copyright law, the author of the work is the physical (natural) person who created the work (a 'romantic author/creator' idea),[74] and they would also normally be the initial owner of the work.[75] As an exception, works created by an employee in the course of employment are owned by the employer, unless there is a contract stating otherwise.[76] Likewise, under patent law, the inventor is usually the natural person who conceived the invention,[77] and they are also the initial owner of the invention. Like copyright law, in cases involving an employment relationship, the employer is automatically the first owner of the invention and, eventually, of the patent.[78]

---

[71] 545 U.S. 913, pp. 934-935 (2005).
[72] 464 U.S. 417, p. 439 (1984).
[73] E.g. Joshua L. Simmons, "Inventions Made for Hire" (2012) 2(1) *New York University Journal of Intellectual Property and Entertainment Law* 1-50, pp. 43-47.

*DABUS* triggered an interesting inquiry concerning IP ownership of AI-generated output. While the High Court of England and Wales confirmed that an AI system cannot be listed as an inventor, it opened up the possibility of listing an *owner of AI* as both the *inventor* and the *owner* of the patent on an AI-generated invention. Given the above similarities of rules governing ownership in patent law and copyright law, it is worth investigating whether the allocation of the initial ownership rule concerning AI-generated output in *DABUS* could be suitable in a copyright law context.[79]

---

[74] E.g. Christopher Aide, "A More Comprehensive Soul: Romantic Conceptions of Authorship and the Copyright Doctrine of Moral Right" (1990) 48(2) *University of Toronto Faculty of Law Review* 211-228.

[75] The exception would the rule relating to employee's works, discussed below. Also, this is different in case of neighboring or related rights recordings, broadcasts or cinematographic films. Under most copyright laws, there is no 'author' of these types of subject matter and the initial owners are those who produced the work (ie record company, broadcaster, film maker). Notably, underlying works (such as music, text) would still have authors. One of the exception is the UK, where the 'author' is defined broadly and includes not only a creator but also a producer of a music recording or a broadcast, as well as a publisher of an edition, see CDPA 1988 s. 9(2).

[76] E.g. CDPA 1988 (UK), s. 11(2); Copyright Act 1986 (Australia), s. 35(6).

[77] E.g. UK Patent Act 1977, s. 7(3), ('actual deviser'); for further discussion see Andrew Stewart et al., *Intellectual Property in Australia* (New York: Lexis Nexis, 2018), p. 469.

[78] Stewart et al., *supra* n. 77, pp. 473-481.

[79] As indicated in the introduction, we will focus on initial ownership only, and leave the question of authorship outside the scope of this paper.

## 3.2  *DABUS* and the (potential) 'owner of AI' rule

In *DABUS*, the High Court of England and Wales concluded that, under the Patent Act 1977 (UK), the inventor must be a natural person.[80] What is more relevant for the purpose of this article is the court's suggestion that, in cases involving AI-generated inventions, the AI owner should be the owner of the invention. According to Smith J,

> (…)there is a general rule that the owner of a thing is owner of the fruits of that thing. Thus, the owner of a fruit tree will generally own the fruit produced by that tree.[81]

Smith J suggested that this analogy applies in considering ownership of AI-generated inventions. As a result, the court concluded that the owner of the DABUS system should own the system's outputs.[82] This may be the first court decision indicating that the AI owner should also be the owner of IP rights over the AI-generated output. We refer to this ownership allocation rule as the *AI-owner approach*.

Additionally, it is worth noting that Thaler, who was the patent applicant, was not a mere 'owner'; he was also the person who created this machine, patented it, possessed it, and used it to generate an invention claimed in the subject's patent application. Smith J was aware of this and stated that Thaler could "rely on this ownership and control of DABUS" to claim his entitlement of patent.[83] He then made a further reservation:

---

[80]  *DABUS*, para. 35.
[81]  *DABUS*, para. 49(3)(a).
[82]  *DABUS*, para. 49(2).
[83]  *DABUS*, para. 49(2).

> I proceed on the basis that Dr Thaler is the only person involved in the ownership and operation of DABUS. If – contrary to my conclusion – ownership or something like it were sufficient to effect a transfer of the invention or the right to apply for a patent, it would be necessary to articulate clearly what forms of ownership and/or control would suffice. These are not matters that I need to consider in this judgment.[84]

This suggests that while the court is generally ready to accept that the AI owner is the owner of the patent in AI-generated outputs, further discussion is needed on what role control of AI plays in allocating ownership over AI-generated outputs. We address this question later in the paper.[85] Below, we first assess the strengths and weaknesses of this AI-owner approach in the area of patent law and then examine its applicability in copyright law.

### 3.3 The AI-owner approach in a patent law context

The ownership allocation rule in *DABUS*, or similar rules, has appeared in the patent law literature.[86] For instance, Ryan Abbott, who is one of coordinators of DABUS litigation around the world, strongly supports this ownership allocation option.[87] According to Abbott, the main reason for allocating initial ownership of AI-generated inventions to the owner of AI is "because this is most consistent with the rules governing ownership of property and it would most incentivise innovation".[88] Namely, allocating ownership to the AI owner as opposed to the user would arguably incentivise the provision of access to the AI and, thus, more

---

[84] *DABUS*, n. 34.
[85] See section 3.4.2 below.
[86] See Abbott, *supra* n. 14, pp. 1114-1117; for other overview of other ownership allocation proposals see Pearlman, *supra* n. 7, pp. 25-30.
[87] Abbott, *supra* n. 14, pp. 1114-1117.
[88] *Ibid.*, pp. 1113-1114.

*Matulionyte and Lee*                                                                 27innovation. For instance, IBM's Watson program, which was initially designed to compete on the game show *Jeopardy!* and to invent new food recipes, was subsequently made available to different software application providers.[89] This enabled them to create services with Watson's capabilities, and Watson is now assisting with financial planning, the development of treatment plans for cancer patients, the identification of potential research study participants, distinguishing genetic profiles that might respond well to certain drugs, and acting as a personal travel concierge.[90] If Watson invented something while being used by other users and those users owned the invention by default, IBM would be disincentivised to give access to Watson to other users. If, however, users wanted to own Watson-generated inventions, this would require an agreement and possibly a fee given to IBM.[91] Unsurprisingly, IBM has expressed its support for the AI-owner approach in patent law.[92]

On the other hand, this AI-owner approach might raise a few issues. For example, what if the owner of AI does not contribute anything substantial to the invention – should they still own it? If DABUS AI was sold to a company that uses it to make inventions but does not contribute to these inventions in any substantial way, is it reasonable that a (new) DABUS owner also owns patents on inventions generated by DABUS? While some might question the viability of this result,[93] we suggest that such an outcome is reasonable. Ownership does not require any substantial contribution; rather, it is about the amount of investment. IP generated in the course of employment and IP assignment are both notable examples. If you invested in ownership, you should own the outputs. If you bought a garden with apple trees, you own the apples, even if you did not invest in planting and taking care of the garden initially. A similar rationale underlies rules on copyright ownership of computer-generated works in the CDPA 1988. Scholars have argued that computer-generated works under CDPA 1988 are nearer to entrepreneurial works than to authorial works because their production does not involve human creativity.[94] Authorial works are protected in copyright law due to the originality and creativity contributed by their authors, whereas entrepreneurial works are protected to incentivise investment in making specific works available to the public.[95] If we follow this line of reasoning, it would be reasonable to assign ownership of a computer-generated work to the person investing in the production of the work rather than the person contributing originality to the work.

---

[89] *Ibid.*, pp. 1089-1090.
[90] *Ibid.*, p. 1091.
[91] *Ibid.*, p. 1115.
[92] USPTO, *supra* n. 8, p. 7.
[93] Abbott, *supra* n. 14, p. 1116.

Secondly, if the owner of the AI is the owner of the AI-generated invention, AI developers may prefer licensing their AI over selling it.[96] This possible development may be reinforced by the practice of the information technology (IT) industry, where software is seldom 'sold' or, using IP law terminology, assigned. Instead, software is normally licensed under the terms of sole, exclusive, or non-exclusive license.[97] If such practice remains in the AI industry, developers would in most cases remain the owners of AI even in cases where they give exclusive or sole licenses to users who then actually control the AI system and plan to commercially exploit the resulting AI-generated invention. We discuss this issue in a subsequent section.[98]

---

[94] Bond and Blair, *supra* n. 31, p. 423; Lionel Bently and Brad Sherman, *Intellectual Property Law* (Oxford University Press, 4th ed., 2014), p. 117; Dornis, *supra* n. 7, pp. 44-46; Lambert, *supra* n. 10, pp. 13, 18; Maggiore, *supra* n. 7, p. 398.

[95] Richard Arnold, "Content Copyrights and Signal Copyrights: The Case for a Rational Scheme of Protection" (2011) 1(3) *Queen Mary Journal of Intellectual Property* 272-279, p. 277; Lee, *supra* n. 24, pp. 184-186.

[96] See Abbott, *supra* n. 14, p. 1116.

[97] For the definitions of these see e.g. Stewart et al., *supra* n. 77, pp. 848-855.

## 3.4 AI-owner approach in copyright law context

In the context of copyright, allocating ownership of AI-generated creative works to the AI owner is an innovative idea. It is different from other options, such as the AI software developer or AI software user as an owner of AI-generated works, as discussed in section 2. In some cases, the software developer, user, and owner will coincide, and the issue of how to distinguish them will not arise. Thaler, who developed, owned, and used DABUS, is a classic example in patent law. In copyright law, it is also possible that a company develops an AI and uses it to generate works, thus giving them copyright in those works (assuming that the AI-generated works are protected under copyright in the first place). The *Tencent* case in China discussed above is an example of such a scenario.[99] In other cases, the developer, user, and owner might be three (or more) different parties. For instance, a software company (developer) may develop an AI system, sell it to another company (owner) which then licenses it to consumers (users) who use it to generate works. The debate over whether the AI developer, AI user, or the AI owner should be the owner of the AI-generated works makes sense when these three roles are played by different parties. In this section, we will identify the advantages and challenges of the proposal of implementing the AI owner rule in the copyright context and analyse whether this approach is more viable than previously discussed ownership allocation options.

---

[98] See section 3.4.2 below.
[99] See section 2.1.2 above.

*3.4.1    Advantages*

Nominating the AI owner as the owner of the AI-generated works has several advantages. First, it might be difficult to determine who – the AI developer, AI user or a third party, such as data trainer or provider – made sufficient contribution (or 'necessary arrangements' under the CDPA 1988 in the UK) to the final output of the AI.[100] If ownership is allocated to the AI owner, there is no need to determine who made what input or whose input (developer's or user's) is the most indispensable to the AI generative process. The allocation is much more straightforward: if you own the AI, you own its AI-generated outputs. This ensures more legal certainty and foreseeability. It also reflects the general principles of property law: if you own an apple tree, you own the apples.

Second, both the software-developer-as-the-owner and the user-as-the-owner approaches emphasise the essential contribution made by the parties in the creation of the output. Therefore, both approaches face the dilemma of potentially leaving AI-generated works in the public domain if neither the developer nor the user has made sufficient or direct contributions to the final output.[101] If the AI owner is considered to be the owner of the AI-generated outputs, such a problem is unlikely to emerge.[102] If we agree that IP protection for AI-generated works is a desirable policy, it is clear that the AI owner rule can achieve this policy goal with fewer transaction costs than the other two approaches.

Third, co-ownership situations are less likely to arise or are likely to be less problematic if the AI owner rule is applied. If ownership of AI-generated outputs is allocated to a person who contributed to the final outputs, there might be situations where both the AI developer and AI user provided sufficient contributions and, thus, qualify as co-owners.[103] In such situations, the exercise of rights might become difficult and costly, especially in situations without a pre-established commercial relationship. The significant transaction costs of such co-ownership will make the copyright of AI-generated work less valuable and, consequently, its consumption may be below an socially optimal level.[104] For instance, if a musician uses Google's AI system Magenta to generate music[105] and both Google and the musician are recognised as co-owners of the AI-generated music, the exercise of rights to the song would become difficult. The musician might need Google's permission to license or transfer the rights, and *vice versa*.[106] If the AI owner is given ownership over the AI-generated outputs, a co-ownership situation is less likely to emerge. Most often, AI is developed by a single company. Even if the AI is developed by several persons or companies and, therefore, several people co-own the AI,[107] it would be easier for them to manage the co-ownership relationship since they already have a working relationship. There might be instances where several AI modules, owned by different entities, are used to produce the final output (eg AI-enabled software writing the text and software editing the output), which may lead to co-ownership situations without pre-existing relationship between owners. However, it is to be seen how frequently these complex situations arise and whether they could be tackled through contractual arrangements discussed below.

---

[100]   See discussion above.
[101]   Ginsburg and Budiardjo, *supra* n. 12, pp. 533-445.
[102]   Certainly, there might be disputes as to the ownership of AI system, especially if it was developed outside employment relationship or if data used to train system was not properly acquired or licensed. However, the ownership of AI falls outside the scope of this paper.

---

[103] E.g. Ginsburg and Budiardjo, *supra* n. 12, p. 440.
[104] Jyh-An Lee, "Copyright Divisibility and the Anticommons" (2016) 32(1) *American University International Law Review* 117-164, pp. 124-130.
[105] See https://magenta.tensorflow.org/ (accessed 20 April 2021).
[106] E.g. Stewart et al., *supra* n. 77, p. 197.
[107] This might happen e.g. when a few or a group of individuals develop AI system outside an employment relationship, or where a few companies or organizations are collaborating to develop the AI system.

Finally, as discussed above, if owners of the AI are allocated ownership over the AI-generated outputs, they would have incentives to make the AI system available to users: regardless of the contributions of the users, the AI owners would own the outputs of the AI. Arguably, IBM would not have given users access to the AI Watson if the company had not been able to claim ownership over the outputs that Watson generated.[108] At the same time, other ownership arrangements could be made if users are not satisfied with this default rule. For instance, if the user wants to own AI-generated outputs, they might agree to pay (higher) licensing fees, which would compensate for the investments that the AI owner made in developing or acquiring the AI system. Similarly, if AI developer is not interested in keeping ownership over AI-generated outputs (as well as any responsibilities that it may cause), they may contractually assign their rights to the outputs to the AI user/licensee.

### 3.4.2 Challenges

At the same time, the AI-owner rule seems unreasonable in situations where an AI system is licensed to another party who uses it to produce commercially valuable outputs over which the party would expect to have exclusive control. For example, if an IT company develops an AI system that produces media articles and licenses it to a media company, the latter would expect that they own, or at least can exclusively use, the media articles autonomously produced by the AI system. If they do not enjoy such rights over the AI outputs and thus are restricted from using them in their commercial practice, they would be discouraged from using the AI system in the first place. Similar concerns were revealed in the *Feilin* case in which the court held that legal rights, if any, should be assigned to the user of the software instead of its owner or developer.[109] This issue becomes even more obvious when an AI system is licensed to end users who have no negotiating power to influence the terms of license. For instance, the user of the AI system Magenta might invest time and effort in making Magenta generate a song they like, and it might seem unreasonable if the user does not own, or at least cannot exploit, the work according to their business plans. These end users do not have sufficient bargaining power to negotiate terms that are different from the standard terms of use provided by the platform, especially if they are able to access the system for free. Therefore, considering the business practicability of such users, the AI-owner rule does not seem to be an ideal option.

---

[108] Abbott, *supra* n. 14, p. 1115.

   This may be the reason that in *DABUS*, Smith J held that it was reasonable to allocate ownership of the AI-generated invention to Thaler, the AI owner, assuming that he was the only person who both owned and controlled the AI when it produced the invention.[110] Smith J also reasoned that Thaler could claim ownership over the AI-generated invention because he controlled the DABUS AI system. Smith J may have been considering a situation wherein the AI system was licenced to a party that was not the AI owner and that party planned to commercially exploit the AI-generated invention. To put it differently, Smith J implied that Thaler might be the owner of the invention generated by DABUS because he had not licenced the AI system to another party to generate new inventions. This understanding of *DABUS* leads to two questions concerning the AI-owner approach. First, should this approach require that the AI owner have legal and factual control over the AI in order to own the AI-generated output? Second, is the AI-owner rule still feasible given the possible scenario wherein an AI system is licensed to another party for commercial use?

---

[109] See section 2.2 above.
[110] *DABUS*, n. 34.

As to the first question, we believe adding the control factor to the AI-owner rule does not help. The infeasibility of adding this is obvious. When AI is used by a licensee (i.e., one party owns the AI and the other party has control over the AI), neither party would be able to meet the requirements of the own-and-control test. In other words, the ownership allocation rule will fail to identify the appropriate owner, and the AI-generated output would, therefore, be in the public domain. If we believe that AI-generated works should be protected by IP, putting them in the public domain is certainly not desirable.

As to the second question, we trust that the AI-owner rule is still a reasonable default rule for ownership allocation because, in addition to the advantages identified in section 3.4.1, the market itself can properly regulate the situation where the licensee is the primary user of the AI. All default property allocation rules only provide a common parameter for various dimensions concerning the subject property, and these rules are always subject to contractual adjustment by private parties. If AI owners do not provide users with sufficient rights to use or commercialise the content they produce using an AI system, users may stop using the system and shift to other AI software vendors. Consequently, such AI owners will likely be pressured by market competition to set acceptable licensing terms.

In summary, an ideal default rule for initial ownership allocation can, under normal conditions, reduce transaction costs between parties.[111] While the rule is desirable for most parties, other parties can easily contract around it with low transaction costs.[112] Based on this understanding, we argue that the AI-owner approach is a feasible option for ownership allocation because it provides significant legal certainty in ownership and leads to transaction costs lower than those brought by the person-who-made-necessary-arragments (AI-software-developer or the AI-software-user) approach. Although, like all default property allocation rules, the AI-owner approach cannot address property interests in every social relationship, it can be adjusted by private ordering through contractual arrangements.

---

[111] E.g. Dan L. Burk and Brett H. McDonnell, "The Goldilocks Hypothesis: Balancing Intellectual Property Rights at the Boundary of the Firm" (2007) 2 *University of Illinois Law Review* 575-636, p. 618.

## 4 Conclusion

The goal of this paper was to revisit the discussion on copyright ownership of AI-generated content and to provide an original analysis of whether the AI-owner rule, as recently proposed in *DABUS* by the High Court of England and Wales, could be a more viable ownership allocation option than other approaches proposed so far (mainly, necessariy arrangement test who allocates ownership either to software-developer or software-user). This paper has demonstrated that while the AI-owners rule may not properly address the end user's commercial considerations in certain situations, this issue can usually be resolved by market competition and private ordering. More importantly, the AI-owner rule provides legal certainty and generates lower transaction costs than the previously proposed approaches.

---

[112] Richard S. Murphy, "Property Rights in Personal Information: An Economic Defense of Privacy" (1996) 84(7) *Georgetown Law Journal* 2381-2418, p. 2412.